\documentclass[sigconf]{acmart}
\AtBeginDocument{%
  \providecommand\BibTeX{{%
    \normalfont B\kern-0.5em{\scshape i\kern-0.25em b}\kern-0.8em\TeX}}}

\setcopyright{rightsretained}
\copyrightyear{2024}
\acmYear{2024}
\acmDOI{}

\acmConference[A3DE@HRI'24]{Assistive Applications, Accessibility, and Disability Ethics workshop at the HRI Conference}{March 15, 2024}{Boulder, CO}
\acmBooktitle{A3DE workshop at HRI} 
\acmISBN{}

\begin{document}
\title[Love, Joy, and Autism Robots: A Metareview and Provocatype]{Love, Joy, and Autism Robots:\\A Metareview and Provocatype}

\author{Andrew Hundt}
\email{ahundt@cmu.edu}
\orcid{0000-0003-2023-1810}
\affiliation{
  \institution{Carnegie Mellon University}
  \streetaddress{5000 Forbes Ave}
  \city{Pittsburgh}
  \state{PA}
  \country{USA}
}

\author{Gabrielle Ohlson}
\email{gohlson@cs.cmu.edu}
\orcid{0009-0008-5122-8028}
\affiliation{
  \institution{Carnegie Mellon University}
  \streetaddress{5000 Forbes Ave}
  \city{Pittsburgh}
  \state{PA}
  \country{USA}
}

\author{Pieter Wolfert}
\email{pieter.wolfert@donders.ru.nl}
\orcid{0000-0002-7420-7181}
\affiliation{%
  \institution{Radboud University}
  \streetaddress{Houtlaan 4}
  \city{Nijmegen}
  \country{The Netherlands}}

\author{Lux Miranda}
\email{lux.miranda@it.uu.se}
\affiliation{%
  \institution{Uppsala University}
  \country{Sweden}}

\author{Sophia Zhu}
\email{sophiazh@andrew.cmu.edu}
 \orcid{0009-0007-6221-4847}
\affiliation{
  \institution{Carnegie Mellon University}
  \streetaddress{5000 Forbes Ave}
  \city{Pittsburgh}
  \state{PA}
  \country{USA}
}

\author{Katie Winkle}
\email{katie.winkle@it.uu.se}
\orcid{0000-0002-3309-3552}
\affiliation{%
  \institution{Uppsala University}
  \country{Sweden}}

\renewcommand{\shortauthors}{Hundt and Ohlson, et al.}

\begin{abstract}
Previous work has observed how Neurodivergence is often harmfully pathologized in Human-Computer Interaction (HCI) and Human-Robot interaction (HRI) research~\cite{williams2021misfit,miranda_examining_2023,anonymous2020epistemicviolencedisability,speil2019autisticagencychildrenreview}. 
We conduct a review of autism robot reviews and find the dominant research direction is Autistic people's \textit{second to lowest} (24 of 25) research priority~\cite{cage2024autisticpriorities}: interventions and treatments purporting to `help' neurodivergent individuals to conform to neurotypical social norms, become better behaved, improve social and emotional skills, and otherwise `fix' us-- rarely prioritizing the internal experiences~\cite{ratto2023innerexperienceautism} that might lead to such differences. 
Furthermore, a growing body of evidence indicates many of the most popular current approaches risk inflicting lasting trauma and damage on Autistic people. 
We draw on the principles and findings of the latest Autism research, Feminist HRI~\cite{winkle2023feministhri} and Robotics~\cite{hundt2022robots_enact,hundt2023equitable-agile-ai-robotics-draft} to imagine a role reversal, analyze the implications, then conclude with actionable guidance on Autistic-led scientific methods and research directions.

\end{abstract}

\received{16 February 2024}
\received[accepted]{23 February 2024}
\received[camera ready]{7 March 2024}

\maketitle

\section{Introduction and Background}
Researchers have created a plethora of robots to `correct issues' in Autistic people (Sec \ref{subsec:metareview}), and this motivated our joint team of neurotypical and neurodivergent researchers to ask: 
In place of designing robots for `Autism Spectrum Disorder (ASD)', how might we design robots to address the understudied phenomenon of `\textit{Neurotypical Spectrum Disorder (NSD)}'\cite{neurotypical_spectrum_disorder}?
We ask this question as a \textit{provocatype}, a creative design provocation as per Bowles' Future Ethics~\cite[p 24]{bowles2018future}, to elicit conversations, recognise the lived and embodied experiences of neurodivergent individuals, and to improve research directions.
We introduce key concepts, review Autism robot reviews and Autism research (Sec. \ref{sec:autism_research_and_hri_impacts}), Autistic research priorities (Sec. \ref{subsec:autistic_priorities}), return to our Feminist HRI~\cite{winkle2023feministhri,hundt2022robots_enact,hundt2023equitable-agile-ai-robotics-draft} provocatype (Sec. \ref{subsec:neurotypical_spectrum_disorder}, \ref{subsec:robot_provocatype_method_and_reflections}), then conclude with scientific guidance (Sec. \ref{subsec:recommendations_and_guidance}, \ref{sec:conclusion}).

\subsection{Neurodiversity and Disability Terminology}
\textit{Neurodiversity} can be defined as the full range of peoples' neurotypes that exist in the world, including \textit{nondisabled} (able-bodied) people; it is analogous to the biodiversity of living organisms, but for brains and \textit{bodyminds} (body and mind as one). 
The term for individuals with neurologically-related disabilities is \textit{neurodivergent}, while nondisabled people and non-cognitively Disabled people can be described as \textit{neurotypical}.
\citet{berghs2016implications} outlines the medical, social~\cite{union1976union}, human rights~\cite{degener2016human}, and CDS~\cite{milton2023criticalautismstudieshandbook} models of disability, which are crucial to research designs.
Ashley Shew's excellent book \textit{Against Technoableism}~\cite{shew2023againsttechnoableism} briefly introduces key terminology, concepts, the Autistic community, and analyzes both disability robots and the implications of \textit{technoableism}, defined as follows:
\begin{quote}
\setlength{\leftskip}{-0.3cm}
\setlength{\rightskip}{-0.3cm}
\textit{``Technoableism is a belief in the technology that considers the \textit{elimination of disability} as a good thing, something we should strive for. 
It's a classic form of ableism--- bias against disabled people, bias in favor of non-disabled ways of life. Technoableism is the use of technologies to reassert those biases, often under the guise of empowerment.'' ---~Shew~\cite{shew2023againsttechnoableism}} %
\end{quote}
\subsection{History}
Researchers should be familiar with: Autism's initial origins in 1911~\cite{Evans2013}, the 1925 Autism description by Jewish psychiatrist Grunya Efimovna Sukhareva~\cite{ssucharewa1996first};
the Nazi T-4 campaign to systematically exterminate Autistic people with high support needs (and Disabled people generally) and complicit physician Hans Asperger~\cite{czech2023aspergerresponse,czech2018asperger} (who never cited Sukhareva); the historical abuse of Autistic people in research studies~\cite{bottema2023scientificantiableism,silberman2016neurotribes}, \textit{e.g.} in ABA (Sec. \ref{subsubsec:aba}); and today's ableism crisis in Autism research~\cite{botha2022autismresearchcrisis,Botha2021}. 
Positive events include knowledge from and about Autistic people's lives~\cite{omeiza2024,price2022unmasking,obrien2023}; the Disability Rights~\cite{charlton1998nothing,silberman2016neurotribes}, Disability Justice~\cite{invalid2019skin,schalk2022black,omeiza2024}, Autism Rights~\cite{silberman2016neurotribes}, and Neurodiversity~\cite{silberman2016neurotribes} movements; the Autistic Self Advocacy Network (ASAN)~\cite{asan}; and Autistic Doctors International~\cite{autisticdoctors}. 
See \textit{Neurotribes}~\cite{silberman2016neurotribes} by Steve Silberman for details.

\section{Autism Research and HRI Impacts}
\label{sec:autism_research_and_hri_impacts}
Next, we cover research findings that impact HRI research and our metareview.
\citet{Williams_2018}\cite{anonymous2020epistemicviolencedisability}, \citet{jackson2022disabilitydongle}, \citet{speil2019autisticagencychildrenreview}, and \citet{keyes2020automating} are leading innovative and impactful perspectives on technologically-based Autism research.
For example, the dominant approach to Autism robotics research risks creating \textit{Disability Dongles}~\cite{jackson2022disabilitydongle}, or devices made without the population the research is about, leading to outcomes unfit for the intended purpose.
\subsection{Review of Autism Robot Reviews}
\label{subsec:metareview}
We consider 12 reviews related to Autism robots since 2020~\cite{kouroupa2022autismrobotreview,scassellati2012robotautismreview,saleh2021autismrobotreview,Sani-Bozkurt2021autismsocialrobotreview,salimi2021autismrobotreview,Alabdulkareem2022autismrobotreview,cano2021affectiveautismrobotreview,abuamara2021autismrobotreview,Puglisi2022,kewalramani2023scoping,kohli2022robot,ragno2023reviewsocialrobotsinhealthcare}, highlighting standout papers and important findings.
Exceptionally robust systematic Autism robot reviews are \citet{wallbridge2024autismrobotreview}, \citet{salimi2021autismrobotreview}, and \citet{kewalramani2023scoping}. 
\citet{wallbridge2024autismrobotreview} is coproduced with Autistic people and the absolute strongest review; covering the process of familiarizing Autistic people with robotic systems while noting withdrawals from studies and areas of limited evidence.
\citet{salimi2021autismrobotreview} observe significant faults in prevailing Autism research methods, stating that, ``The main limitations of current studies [on Autism robots are the] shortage of [Randomly Controlled Trials (RCTs)], low power, and bias.''
\citet{kewalramani2023scoping} noted that more high-quality studies are needed, recognizes the agency of Autistic children, and minimizes the use of deficit language.
\citet{abuamara2021autismrobotreview} astutely observes potential benefits of low social risk, enjoyable, and predictable robot interactions.
Some reviews~\cite{Puglisi2022,ragno2023reviewsocialrobotsinhealthcare,bartl-pokorny2021autismrobotreview} cover physical hardware.

Despite these strengths, we observe critical shortcomings in most reviews.
With \citet{wallbridge2024autismrobotreview} excepted, no review seriously considers the research priorities Autistic people set for themselves (Sec. \ref{subsec:autistic_priorities}). 
Some reviews accurately note the lack of high-quality efficacy evidence~\cite{abuamara2021autismrobotreview,kohli2022robot,salimi2021autismrobotreview,kewalramani2023scoping} for these devices, but more inaccurately overclaim efficacy~\cite{kewalramani2023scoping,damianidou2020autismrobotsreviewsocialinteractions,Puglisi2022,ragno2023reviewsocialrobotsinhealthcare,saleh2021autismrobotreview,Alabdulkareem2022autismrobotreview}.
Three use inclusive language \cite{wallbridge2024autismrobotreview,kewalramani2023scoping,ragno2023reviewsocialrobotsinhealthcare} (Sec. \ref{subsubsec:language_and_feedback}), but all others use harmful deficit language \cite{kouroupa2022autismrobotreview,scassellati2012robotautismreview,saleh2021autismrobotreview,Sani-Bozkurt2021autismsocialrobotreview,salimi2021autismrobotreview,Alabdulkareem2022autismrobotreview,cano2021affectiveautismrobotreview,abuamara2021autismrobotreview,Puglisi2022,kohli2022robot} and generally tend to promote research well-known to risk harm to Autistic people, without discussing or measuring those harms. 
For example, camouflaging via eye contact (Sec. \ref{subsubsec:camouflaging_masking}) can be very uncomfortable or even painful for many Autistic people~\cite{trevisan2017adults,engelbrecht2024embraceautism}, who rate the task as undesirable~\cite{chazin2024centering}, which should should preclude it from most HSR, yet its the most common application for these robots~\cite{damianidou2020autismrobotsreviewsocialinteractions} and many reviews promote it~\cite{Alabdulkareem2022autismrobotreview,damianidou2020autismrobotsreviewsocialinteractions,abuamara2021autismrobotreview,kohli2022robot,ragno2023reviewsocialrobotsinhealthcare}.
Future research and reviews should be based on up-to-date Autism research.%

\subsection{Current Autism Research}
\label{subsec:current_autism_research}
Next we cover current Autism research findings that provide opportunities for actionable Autism Robot and HRI improvements.
\subsubsection{Language and Feedback}\label{subsubsec:language_and_feedback} Avoid ableist language~\cite{bottema-beutel2021avoidingableist,natri2023anti,bottema2023scientificantiableism} for positive, scientifically-backed research communication~\cite{bottemabeutel2023antiableismscientific,natri2023anti}, which is Autistic priority 5 of 25~\cite{cage2024autisticpriorities} %
(Sec. \ref{subsec:autistic_priorities}). Onboard uncomfortable feedback~\cite{hundt2023equitable-agile-ai-robotics-draft} and avoid derailing it~\cite{ahmed2021complaint,jackson2022disabilitydongle,derailing2018}.

\subsubsection{Autistic Strengths} \citet{cope2022strengths} find that Autistic strengths in the workplace tend to include ``cognitive advantages such as superior creativity, focus, and memory; increased efficiency and personal qualities such as honesty and dedication; and the ability to offer a unique autism-specific perspective.'' 
Our research priorities should not only consider, but actively seek out and emphasize common Autistic strengths in Autistic populations.
\subsubsection{Double Empathy Problem} \label{subsubsec:double_empathy_problem}
\citet{milton2012doubleempathy} is a landmark paper proposing the theoretical framework for the two-way Double Empathy Problem---namely that Autistic-to-Autistic and neurotypical-to-neurotypical communication are each typically effective, but neurotypical-to-Autistic communication frequently involves mis-communication---~elegantly demonstrated by subsequent papers \citet{crompton2020autisticpeertopeer} and \citet{jones2023doubleempathyinteractions} who test communication among Autistic, neurotypical, and mixed groups.
Another landmark paper, \citet{sasson2017neurotypical}, concludes that \textit{Neurotypical Peers are Less Willing to Interact with Those with Autism based on Thin Slice Judgments}.
Non-Autistic people, \textit{e.g.} researchers, clinicians, and parents, tend to demonstrate a \textit{lack of reciprocity}~\cite{gernsbacher2021reciprocity,gernsbacher2006behavior}, or symmetrical exchange as equals where neither Autistic nor non-Autistic people have a dominant position. 
This merits interventions~\cite{pukki2022autisticperspectives} and addressing within research teams. 
\subsubsection{Triple Empathy Problem---Negative Communication Impacts on Autistic People in Healthcare Settings}
\label{subsubsec:triple_empathy_problem}
\citet{doherty2023autistichealthcarebarriers} and \citet{shaw2023tripleempathy} crucially demonstrate how gaps in healthcare providers' methods represent a novel postulated \textit{triple empathy problem} that denies Autistic people access to essential healthcare support. 
Research and Development (R\&D) protocols for Autism robots as health interventions need to be updated accordingly, as per \citet{hundt2023equitable-agile-ai-robotics-draft}.
We recommend \citet{doherty2023autistic}'s \textbf{SPACE: `Sensory needs, Predictability, Acceptance, Communication and Empathy'} framework for actionable guidance to address core Autistic needs and mitigate negative outcomes.
\subsubsection{Autistic Camouflaging (Masking)}\label{subsubsec:camouflaging_masking} Autistic people tend to seek genuine connections with others, but in adverse or unsafe environments, many Autistic people must hide Autisic traits to appear `more neurotypical'---a process known as \textit{camouflaging} or \textit{masking}~\cite{pearson2021conceptual,price2022unmasking}. 
Masking is not necessarily a choice~\cite{pearson2021conceptual} and has known mental health impacts~\cite{hull2021social,perry2022understanding}. 
Human Subjects Research (HSR) and clinical research require principles of \textit{beneficence} and \textit{nonmaleficence} be respected, so the known adverse impacts~\cite{hull2021social,perry2022understanding} preclude most robotics and HRI research from encouraging masking or camouflaging, broadly construed (\textit{e.g.} eye contact~\cite{trevisan2017adults} training), including through play-based and naturalistic methods~\cite{pukki2022autisticperspectives}.
\subsubsection{Diversity of Autistic People and Support Needs} Describe specific support needs and include Autistic people who are diagnosed and self-assessed~\cite{mcdonald2020autism}, nonspeaking, children~\cite{chazin2024centering}, adults, as well as the range of identities with respect to race, gender, sexual orientation, co-occurring disabilities \textit{e.g.} Intellectual Disabilities (ID) and wheelchair users, plus other identities~\cite{pukki2022autisticperspectives}. 
For example, 37\% (95\% CI 28–46\%)\cite{micai2023prevalence} of Autistic people have co-occurring ADHD. Use ASAN accessibility guides~\cite{asan2019accessibleeventplanning,asan2021easyread} and the book \textit{Research Involving Participants with Cognitive Disability and Difference}~\cite{casico2018participantswithcognitivedisability}.
\subsubsection{Autistic Joy}
Autistic Joy~\cite{pellicano2022capabilities,price2022unmasking} comes in countless forms, \textit{e.g.} engaging deeply with our special interests, and stimming, or self-stimulation, as a joyful self-regulating activity. 
Many Autistic people find joy in sharing information, knowledge, and resources (infodumping).
Many respond to others who share personal experiences with ones of their own to relate and build bonds. 
\subsubsection{Applied Behavioral Analysis (ABA)}\label{subsubsec:aba}A controversial~\cite{sandoval2019much,silberman2016neurotribes,ann_memmott_blog,kupferstein2018evidence} range of treatments mainly lacking rigorous empirical evidence, aside from some support for Naturalistic (NDBI) and technology based interventions~\cite{sandbank2020project,sandbank2023autismmetaanalysis}. 
Adverse (negative) events are largely unmonitored and may be common~\cite{sandbank2023autismmetaanalysis,bottema2021adverse,bottema2023autisminterventionissues,bottema2023autisticyouthsystematicinterventionevaluation}.
A low Autistic priority at 24 of 25 (Sec. \ref{subsec:autistic_priorities}).
We suggest research alternatives next.
\subsection{Autistic Priorities for Autism Research}
\label{subsec:autistic_priorities}
Next, we highlight Autistic-led priority shifts~\cite{roche2021research,pellicano2014autism,aarc_focus_groups,aarc_draft_priorities,pellicano2014future,van_den_bosch_report,chazin2024centering} needed for Autism Research. \citet{pukki2022autisticperspectives} is leading work.
Autistic people from Scotland ranked their research priorities across 25 topics in \citet{cage2024autisticpriorities}. The \textbf{top six Autistic priorities} were: (1) Mental health and well-being~\cite{najeeb2024autisticwellbeing}; (2) identifying or diagnosing Autistic people, including post-diagnosis support; (3) Services and support across the lifespan, including social care and healthcare; (4) knowledge and attitudes towards Autistic people and how we view and talk about Autistic people; (5) issues that impact Autistic women; and (6) employment. Autistic Joy is ranked 16 of 25. \textbf{Autistic people rank Treatments and Interventions as their \textit{second to lowest} (rank 24 of 25) research priority}, encompassing Applied Behavioural Analysis (ABA), Positive Behaviour Support (PBS), low arousal approach, social skills training, etc.~\cite{cage2024autisticpriorities}
These findings, together with current research (Sec. \ref{subsec:current_autism_research}), constitute substantial evidence that Autism Robot research should adapt to meet Autistic people's actual priorities and needs.

\section{Methods and Discussion}
\label{sec:methods_and_discussion}
Next, we summarize Neurotypical Spectrum disorder, describe our provocatype, then provide reflections and finally recommendations on improving the implementation of Autism Robots.

\textit{\href{https://web.archive.org/web/20220114165409/https://realtalktherapypdx.com/neurotypical-spectrum-disorder/}{Neurotypical Spectrum Disorder (NSD)}}:
\label{subsec:neurotypical_spectrum_disorder}
Visit \citet{neurotypical_spectrum_disorder} to read their \href{https://web.archive.org/web/20220114165409/https://realtalktherapypdx.com/neurotypical-spectrum-disorder/}{DSM-V style NSD criteria} reframing research-backed differences.
For example, 
neurotypical people with NSD, including researchers and clinicians, are \textit{characterized by deficits} in:
(1) understanding and participating in direct communication (verbal, nonverbal, or text), tending to invent implied content when it is not present;
(2) sustaining specific topics of interest for an extended period of time, leading to expertise;
(3)~providing sufficient detail;
(4)~adhesion to moral principles;
(5)~ability to speak up in unjust social and workplace situations;
(6)~social interactions with neurodivergent people;
(7)~avoiding negative assumptions and judgments about harmlessly divergent people and behaviors;
(8)~flexibility about social conformity;
(9)~recognizing neutral neurodivergent faces, voices, and descriptions, with a tendency towards confident misinterpretations as anger, sarcasm, or condescension;
(10)~respecting people they perceive as lower in a social hierarchy and developing a theory of their minds;
(11)~restricted, repetitive need for small talk (\textit{e.g.} the weather).
No one \textit{deficit} is sufficient for an NSD diagnosis, not everyone has NSD, and people with NSD may not have all \textit{deficits}, but together the presence of enough traits at a sufficient degree of intensity can help identify \textit{NSD sufferers}.

\subsection{Robot Provocatype Method and Reflections}
\label{subsec:robot_provocatype_method_and_reflections}
Some people might find the above NSD preface to our \textit{provocatype} upsetting to read. 
This feeling is one Autistic people reading Autism research are deeply familiar with because we must routinely read extremely crass language written about us, without us~\cite{natri2023anti,bottema-beutel2021avoidingableist}.

Furthermore, we consider continuing, pausing, reworking, and winding down (ending) this project to each be serious, actionable options~\cite{hundt2022robots_enact,hundt2023equitable-agile-ai-robotics-draft}.
We have \textit{Theory of Mind (ToM)}~\cite{gernsbacher2019empirical} (we can imagine being in others' shoes), we are \textit{empathetic}~\cite{kimber2023autisticempathy} people, and we are willing to \textit{reciprocate}~\cite{gernsbacher2021reciprocity,gernsbacher2006behavior} (Sec. \ref{subsubsec:double_empathy_problem}). 
We have considered the perspective of a potentially neurotypical reader and/or researcher who might find this NSD approach offensive, \textit{e.g.} by swapping role names in the text~\cite{gernsbacher2021reciprocity,gernsbacher2018more}, and it seems like it could be quite unpleasant.
Thus, we conclude that our hinted NSD \textit{Robot Provocatype}'s risk of harm is too high.

In place of designing a provocatype `social robot for NSD', we decide we \textit{will not} use this opportunity to create assistive robots that address all of the symptoms of NSD that make our current world a difficult place to live in as an Autistic person.
We \textit{will not} start designing a robot to `cure' NSD with repeated exercises to: sustain interests, speak up about moral dilemmas at work, practice empathy, overcome their thin slice judgments of others, and so on.

Instead, we \textit{will} draw on \citet{begel2020lessonslearnedautisticai}, a project for AI-based emotion recognition with a fantastic outcome, specifically, they ended the experiment once they received negative feedback from Autistic people.
\textit{Therefore, our real \textit{Research Provocatype} is to take responsibility and end this experiment before it even starts.
We hope that those working on `robots for Autism' will reflect on when it might be best for them to reciprocate and do the same.}
\subsection{Recommendations and Guidance}
\label{subsec:recommendations_and_guidance}
We seek research that advances the state of empathetic understanding and creates resources that Autistic communities actively request.
We advise researchers to shift research priorities and project designs to address Autistic people's priorities for Autism Research (Sec. \ref{subsec:autistic_priorities}, \cite{pukki2022autisticperspectives,cage2024autisticpriorities}). 
Autism-related robots should be carefully and inexpensively designed to address top Autistic priorities, then be rigorously evaluated against robot-free alternatives \textit{e.g.} with Randomly Controlled Trials (RCT) or high-quality qualitative studies.

Studies need to explicitly follow frameworks like SPACE~\cite{doherty2023autistic} (Sec. \ref{subsubsec:triple_empathy_problem})
to ensure Autistic people's individual needs are supported in system designs and during the research process.
Also respect Autistic agency~\cite{speil2019autisticagencychildrenreview,memmott2023ethics}; 
record assent, consent, and withdrawals~\cite{memmott2023ethics}; and describe adverse (harmful) events, \textit{e.g.} due to research methods~\cite{memmott2023ethics}.
Read \citet{casico2018participantswithcognitivedisability} for \textit{``suggestions for research design, research ethics, and best practices that empower people with cognitive disabilities and differences to participate in research while respecting and managing potential coercion or undue influence.''}
We suggest keeping up with the journal \textit{Autism in Adulthood} and Autistic-led resources~\cite{autisticdoctors,aaspire,asan,thinking_autism_guide, ann_memmott_blog,engelbrecht2024embraceautism,neurodiverse_connection_uk}.

Furthermore, research and design methods matter. \textit{Robots Won't Save Japan}~\cite{wright2023robots} and its review~\cite{hundt2024roboteldercarereview} share insights on assumptions and limitations all roboticists and HRI researchers should consider. Notably, all of the disability-related robots \citet{wright2023robots} evaluated aimed to reduce the workload and burden of support personnel, but actually \textit{added} to it. \citet{hundt2024roboteldercarereview} connects it to Autism research.

Adopt participant-led~\cite{hundt2023equitable-agile-ai-robotics-draft,botha2022autismresearchcrisis}, co-design, participatory, Design Justice~\cite{costanza2020design, winkle2023feministhri}, and/or Equitable Agile~\cite{hundt2023equitable-agile-ai-robotics-draft} methodologies. 
Better frameworks for project governance~\cite{hundt2023equitable-agile-ai-robotics-draft} can improve research and development outcomes.
Project performance can be evaluated with participant-led research scorecards~\cite{hundt2023equitable-agile-ai-robotics-draft,plrc2023scorecards}.
Ground research in, and cite, current Autism research (Sec. \ref{sec:autism_research_and_hri_impacts}), including critiques.
Use inclusive language~\cite{bottema-beutel2021avoidingableist}.
Consider internal experiences~\cite{ratto2023innerexperienceautism}.
Always ensure Autistic adults or Autistic Autism scholars review proposals, experiment designs, and papers; are offered authorship accordingly, and are \textit{paid} whenever possible.
Finally, onboard inconvenient findings and feedback; it could lead to path-breaking outcomes.

\section{Conclusion}
\label{sec:conclusion}
In summary, we laud the best of recent Autism research and reviews, but find the bulk of research on Autism robots ranges from cases where there is substantial room for improvement up through cases where there is a high risk that harm has occurred. %
Finally, we reviewed actionable resources, findings, and methods to meet Autistic needs and research priorities in HRI and robotics.

\bibliographystyle{ACM-Reference-Format}
\bibliography{bibliography}

\appendix

\end{document}